\documentclass[12pt]{article}
\usepackage{a4wide}
\usepackage{amssymb}
\usepackage{xcolor}
\usepackage{hyperref}
\usepackage[normalem]{ulem}

\newcommand{\bea}{\begin{eqnarray}}
\newcommand{\eea}{\end{eqnarray}}

\def\[{\left[}
\def\]{\right]}
\def\({\left(}
\def\){\right)}
\def\nn{\nonumber}
\def\d{{\rm d}}

\begin{document}
{\renewcommand{\thefootnote}{\fnsymbol{footnote}}
\begin{center}
  {\LARGE  Signature change in 2-dimensional black-hole models\\ of loop quantum gravity}\\
  \vspace{1.5em} Martin Bojowald$^1$\footnote{e-mail address: {\tt
      bojowald@gravity.psu.edu}}
  and Suddhasattwa Brahma$^{1,2}$\footnote{e-mail address: {\tt suddhasattwa.brahma@gmail.com}}\\
  \vspace{0.5em}
$^1$  Institute for Gravitation and the Cosmos,\\
  The Pennsylvania State
  University,\\
  104 Davey Lab, University Park, PA 16802, USA\\
\vspace{0.5em}
$^2$ Center for Field Theory and Particle Physics,\\
Fudan University, 200433 Shanghai, China\\
  \vspace{1.5em}
\end{center}
}

\setcounter{footnote}{0}

\begin{abstract}
  Signature change has been identified as a generic consequence of holonomy
  modifications in spherically symmetric models of loop quantum gravity with
  real connections, which includes modified Schwarzschild solutions. Here,
  this result is extended to 2-dimensional dilaton models and to different
  choices of canonical variables, including in particular the
  Callan-Giddings-Harvey-Strominger (CGHS) solution. New obstructions are
  found to coupling matter and to including operator-ordering effects in an
  anomaly-free manner.

\end{abstract}

\section{Introduction}

General covariance is deformed in most of the existing midisuperspace models
of loop quantum gravity (LQG), in the sense that structure functions in the
hypersurface deformation algebroid do not have the classical form
\cite{DeformedRel}. In particular, non-singular signature change is possible
in the presence of holonomy corrections in models based on real connection
variables; for complex connections see
\cite{SphSymmComplex,CosmoComplex}. These results have been derived by
effective \cite{JR,LTBII,SphSymmCov,GowdyCov} and operator methods
\cite{SphSymmOp}, and they appear in a related form in cosmological
perturbation theory \cite{ScalarHol,DeformedCosmo}.

Since the effective signature of space-time determines the form of well-posed
initial or boundary value problems, modified space-time structures that allow
for signature change have an important influence on the causal behavior of
black-hole models in loop quantum gravity. Signature change results from the
presence of holonomy modifications in the theory, which are the same effects
often used to argue that singularities are removed in such models. However,
even if curvature invariants remain bounded, signature change in the
high-curvature region would imply that there is no deterministic evolution
through high curvature. A black-hole model rather different from simple bounce
models is then obtained, with problematic aspects owing to the presence of new
Cauchy horizons \cite{Loss}. While diverging curvature may be avoided, a
serious problem (much like a naked singularity) persists in the sense of a
space-time incompletely determined by initial data.

In order to extend space-time across the Euclidean region, one requires
additional data on the part of its boundary that borders on the future
Lorentzian space-time.  This additional requirement is reminiscent of other
proposals of black-hole models, for instance `stretched horizons', introduced
in the context of black hole complementarity (see, for instance \cite{Stretched}), except that, with signature change, unexpected degrees of freedom are not located at a horizon.  In any case, the detailed analysis of anomaly-free black hole models
in loop quantum gravity points towards a much more subtle non-singular
description of quantum space-time than usually postulated in simplified bounce
models. Just as an outside observer finds the stretched horizon as a membrane
storing and later releasing information in the form of microphysical degrees
of freedom, additional information is encountered once an observer moves into
the future of a Euclidean region embedded in space-time. However, in the case
of black-hole models of loop quantum gravity, there is as yet no microscopic
theory that would restrict or determine possible data around Euclidean
regions. Geometrically, signature change does indeed demonstrate the existence
of novel non-classical geometries in loop quantum gravity, from which
classical space-time emerges in an effective picture \cite{Normal}.

It is important to note that the same modifications of the classical
dynamics that lead to singularity resolution in terms of bounded curvature are
also responsible for signature change. Bounce models based on real connections
without signature change have been described, following \cite{BHInt}. However,
this has been possible only by using classical gauge fixing or other
assumptions about the structure of space-time, and therefore comes at the
expense of not being able to discuss the all-important anomaly
problem. Without a handle on the anomaly problem, it is not clear whether
there is any consistent space-time structure at all. (In some cases, these models have been shown to also manifest signature change once the anomaly problem has been suitably addressed \cite{SphSymmCov}.)

In this paper, we extend the analysis of signature change in modified
Schwarzschild space-times in two ways. First, we use the formalism of
\cite{SphSymmPSM} in order to study signature change for all 2-dimensional
dilaton models with holonomy modifications. Secondly, we incorporate a
canonical transformation that may be used to modify the kinetic term of the
Hamiltonian constraint (quadratic in extrinsic curvature). Although some of
the formal aspects relevant for signature change seem to be absent for some
choices of such canonical transformations, we will show that signature change
is still realized. Finally, we obtain new partial obstructions to anomaly-free
formulations with holonomy modifications when operator effects or matter terms
are included.

\section{2-Dimensional dilaton gravity}

In two space-time dimensions, the general form of dilaton gravity models has
the first-order action \cite{Strobl}
\begin{equation}
 S=-\frac{1}{2G} \int_M \left(\phi{\rm d}\omega+ \frac{1}{2}V(\phi) \epsilon+
 X_a{\rm D}e^a\right)
\end{equation}
for a dyad $e^a$ with volume form $\epsilon$, the dilaton field $\phi$ and a
connection 1-form $\omega$ which appears in the covariant derivative ${\rm
 D}$. The fields $X_a$ are Lagrange multipliers that ensure
torsion-freedom. The potential $V(\phi)$ is an arbitrary function and
characterizes different models.

\subsection{First-order variables and Poisson Sigma model}

After integrating by parts (ignoring boundary terms), we obtain
\begin{equation}
 S=\frac{1}{2G} \int_M \left(e^a\wedge {\rm d}X_a+ \omega {\rm d}\phi+
   X_a\epsilon^a{}_b \omega\wedge e^b+ \frac{1}{2}V(\phi)\epsilon\right)\,.
\end{equation}
As used in Poisson Sigma models \cite{Ikeda,PSM,Strobl}, it is convenient to
collect the fields in two triplets, $X^i=(X^-,X^+,\phi)$ and
$A_i=(e_x^+,e_x^-,\omega_x)$ of canonical fields, and one triplet
$\Lambda_i=(e_t^+,e_t^-,\omega_t)$ of multipliers. The action then takes the
compact form
\begin{equation}
 S=-\frac{1}{2G} \int_M \left(A_i\wedge {\rm d}X^i+ \frac{1}{2}P^{ij}
   A_i\wedge A_j\right)
\end{equation}
with a Poisson tensor
\begin{equation}
 P=\left(\begin{array}{ccc} 0 & -\frac{1}{2}V(\phi) & -X^-\\ \frac{1}{2}V(\phi)
     & 0 &     X^+\\ X^- & -X^+ & 0\end{array}\right)\,.
\end{equation}

The canonical formulation leads to Poisson brackets
\begin{equation}
 \{X^i(x),A_j(y)\} = 2G\delta^i_j\delta(x-y)\,,
\end{equation}
and three first-class constraints
\begin{equation}
 \tilde{C}^i = \frac{1}{2G} \left((X^i)'+ P^{ij}A_j\right)\,.
\end{equation}
The component $\tilde{C}^3$ generates SO(1,1) rotations of the Lorentzian dyad
and connection, while the linear combination $D= A_i\tilde{C}^i=A_i(X^i)'$
generates spatial diffeomorphisms. The remaining independent combination of
$\tilde{C}^+$ and $\tilde{C}^-$ serves as a Hamiltonian constraint.

\subsection{Variables invariant under SO(1,1)}

A comparison with canonical variables commonly used in spherically symmetric
models, based on a symmetry reduced ADM formulation of canonical gravity, is
facilitated by first introducing combinations of the fields invariant under
the SO(1,1) transformations generated by $\tilde{C}^3$. Instead of $X^{\pm}$
and $e_x^{\pm}$, we follow \cite{SphSymmPSM} and work with
\begin{equation}
 X:=\sqrt{X^+X-} \quad\mbox{ and }\quad e:=\sqrt{e_x^+e_x^-}
\end{equation}
and two boost parameters $\alpha$ and $\beta$ such that
\begin{equation}
 X^{\pm} = X \exp(\pm\beta) \quad\mbox{ and }\quad e_x^{\pm}=e
 \exp(\pm\alpha)\,.
\end{equation}

One can turn these into canonical variables by introducing
\begin{equation}
 Q^e=2X\cosh(\alpha-\beta) \quad\mbox{ and }\quad
 Q^{\alpha}=2X\sinh(\alpha-\beta)
\end{equation}
such that
\begin{equation}
 \{Q^e(x),e(y)\}=\{Q^{\alpha}(x),\alpha(y)\}=\{\phi(x),\omega_x(y)\}=
2G\delta(x-y) \,.
\end{equation}
The original variables can be obtained from the canonical ones as
\begin{equation}
 X^{\pm} = \frac{eQ^e\mp Q^{\alpha}}{2e} \exp(\pm\alpha)\,.
\end{equation}
We now have the constraints
\begin{equation} \label{Cpm}
 \tilde{C}^{\pm} = \frac{1}{2G} \left(\left(\frac{eQ^e\mp
       Q^{\alpha}}{2e}\right)' \pm \frac{eQ^e\mp Q^{\alpha}}{2e}
   (\omega_x+\alpha')\pm \frac{1}{2}V(\phi)e\right)\exp(\pm\alpha)
\end{equation}
and
\begin{equation}
 \tilde{C}^3 = \frac{1}{2G} (\phi'+Q^{\alpha})\,.
\end{equation}

\subsection{Transformation to standard variables of spherical symmetry}

The usual connection or extrinsic-curvature variables of spherically symmetric
gravity \cite{SphSymm} are finally obtained by a canonical transformation from
$(Q^e,e;Q^{\alpha},\alpha;\phi,\omega_x)$ to
$(K_{\varphi},E^{\varphi};K_x,E^x;\eta,P^{\eta})$ with
\begin{eqnarray}
 Q^e &=& 2\sqrt{2} (E^x)^{1/4} K_{\varphi} \quad,\quad
 e=\frac{E^{\varphi}}{\sqrt{2}(E^x)^{1/4}} \label{Qee}\\
 Q^{\alpha} &=& P^{\eta} \quad,\quad \alpha=-\eta\\
 \omega_x &=& -\left(K_x+\frac{E^{\varphi}}{2E^x}-\eta'\right) \quad,\quad
 \phi=E^x\,. \label{omegaphi}
\end{eqnarray}
(See Eq.~(42) in \cite{SphSymmPSM}.)  A suitable combination of $\tilde{C}^+$
and $\tilde{C}^-$, written in these variables, takes the usual form of the
Hamiltonian constraint
\begin{eqnarray} \label{HDil}
 H[N] &=& C^+[2^{-1/2} N\phi^{1/4}\exp(-\alpha)]- C^-[2^{-1/2}
 N\phi^{1/4}\exp(\alpha)]\nonumber\\
&=& -\frac{1}{2G} \int {\rm d}x N
\left(\frac{K_{\varphi}^2E^{\varphi}}{\sqrt{E^x}}+ 2 \sqrt{E^x}
  K_xK_{\varphi}- \frac{1}{2} E^{\varphi}V(E^x)\right.\\
 &&\qquad\qquad\qquad \left.-
  \frac{((E^x)')^2}{4E^{\varphi}\sqrt{E^x}}+
  \frac{\sqrt{E^x}(E^x)'(E^{\varphi})'}{(E^{\varphi})^2}-
  \frac{\sqrt{E^x}(E^x)''}{E^{\varphi}}\right) \nonumber
\end{eqnarray}
in spherically symmetric variables for the appropriate dilaton potential,
$V(E^x)=-2/\sqrt{E^x}$; see Eq.~(46) of \cite{SphSymmPSM}. For an arbitrary
potential, the equations generalize the connection formulation of spherically
symmetric canonical gravity to arbitrary dilaton models, including the CGHS
model \cite{CGHS} for constant $V$.

At this point, we can compare the results with \cite{NewVar11}. The expression
for $e$ in (\ref{Qee}) is nothing but (51) in \cite{NewVar11}, $Q^e$ in
(\ref{Qee}) is (57), and $\omega_x$ in (\ref{omegaphi}) is (60), just in
slightly different notations.

The formulation of the CGHS model presented \cite{NewVar11} is different. In
particular, the canonical transformation to connection variables is simpler
than in the spherically symmetric model, and the Hamiltonian constraint has
only a term of the form $K_xK_{\varphi}$ in its kinetic part but no
contribution of $K_{\varphi}^2$. It amounts to renaming the SO(1,1)-invariant
variables by $\omega_x\equiv K_x$, $Q^e\equiv K_{\varphi}$ and $e\equiv
E^{\varphi}$. These are the variables introduced in Eq.~(20) of
\cite{SphSymmPSM}, corresponding to a standard first-order formulation in
variables as used in Poisson Sigma models. Since the Hamiltonian constraint
then has only one term $Q^e\omega_x=K_xK_{\varphi}$ of variables identified
with extrinsic-curvature components, there is no term of the form
$K_{\varphi}^2$ in the resulting Hamiltonian constraint, as is clear from (23)
of \cite{SphSymmPSM} (our (\ref{Cpm})) and rederived in \cite{NewVar11}.

\subsection{Contribution to kinetic terms}
\label{s:Kinetic}

As the examples discussed so far show, the form of the kinetic contribution to
the Hamiltonian constraint is not invariant under canonical
transformations. In general, if one starts with a generic combination
$\alpha K_{\varphi}^2 E^{\varphi}/\sqrt{E^x}+ K_xK_{\varphi}/\sqrt{E^x}$ with
some real number $\alpha$, as it appears in the standard connection
formulation of spherically symmetric canonical gravity, one can always
transform to new canonical variables that amount to setting $\alpha=0$. To do
so, we need a new
\begin{equation}
 \tilde{K}_x=K_x+\alpha K_{\varphi}E^{\varphi}/E^x
\end{equation}
and $\tilde{K}_{\varphi}=K_{\varphi}(E^x)^{\beta}$ with a second parameter
$\beta$. If we also choose $\tilde{E}^{\varphi}=(E^x)^{-\beta} E^{\varphi}$
and leave $E^x=\tilde{E}^x$ unchanged, the tilde variables are canonical
provided that $\beta=\alpha$. The new kinetic term is
$\tilde{K}_x\tilde{K}_{\varphi}$ without a contribution from
$\tilde{K}_{\varphi}^2$. The choice made in \cite{LoopSchwarz} for the CGHS
model amounts to applying such a canonical transformation.

Previous derivations of signature change in spherically symmetric models with
holonomy modifications used both contributions to the kinetic term of the
Hamiltonian constraint, based on (\ref{HDil}). If one of them can be removed
by a canonical transformation, one may wonder whether the same conclusions can
still be drawn. In the present paper, we answer this question in the
affirmative, and also comment on possible operator effects as well as matter
couplings.

\section{Modified Schwarzschild Models}

Spherically symmetric models have already been analyzed in detail
\cite{SphSymmCov}. At an effective level, one can work with different kinds of
constraint algebras, given by the usual hypersurface deformation brackets on
one hand and a partially Abelianized system without structure functions on the
other \cite{LoopSchwarz}. The latter is easier to deal with when one attempts
a full quantization, but as shown in \cite{SphSymmCov}, it obscures the
important question of covariance because it is not obvious that a consistent
set of hypersurface-deformation generators may be recovered in the modified or
quantized system. If one analyzes possible realizations of
hypersurface-deformation brackets in the partially Abelianized system with
quantum modifications, the structure functions turn out to be modified and
signature change is obtained in the same way as in a direct treatment of the
brackets. In \cite{SphSymmCov}, the analysis was done at an effective level;
here we complement it with operator considerations, and further
generalizations.

\subsection{Deformed covariance}

It is possible to modify (\ref{HDil}) while keeping the system first-class. In
spherically symmetric models, the possibility of anomaly-free holonomy
corrections has been analyzed based on a Hamiltonian constraint
\begin{eqnarray} \label{Hf}
 H[N] &=& -\frac{1}{2G} \int {\rm d}x N
\left(\frac{f_1(K_{\varphi})E^{\varphi}}{\sqrt{E^x}}+ 2 \sqrt{E^x}
  K_xf_2(K_{\varphi})+ \frac{E^{\varphi}}{\sqrt{E^x}}\right.\\
&& \left.\qquad\qquad\qquad -
  \frac{((E^x)')^2}{4E^{\varphi}\sqrt{E^x}}+
  \frac{\sqrt{E^x}(E^x)'(E^{\varphi})'}{(E^{\varphi})^2}-
  \frac{\sqrt{E^x}(E^x)''}{E^{\varphi}}\right) \nonumber
\end{eqnarray}
with two free functions $f_1$ and $f_2$ of $K_{\varphi}$. The holonomy
modification function $f(K_{\varphi})$ is typically chosen to be $\sin\(\rho
K_\varphi\)/\rho$, but may be kept more general.  The system is anomaly-free if
\begin{equation} \label{f2f1}
 f_2(K_{\varphi})= \frac{1}{2} \frac{{\rm d}f_1}{{\rm d}K_{\varphi}}
\end{equation}
in which case the classical structure functions are modified by an additional
factor of \cite{JR}
\begin{equation} \label{beta}
 \beta(K_{\varphi}) = \frac{{\rm d}f_2}{{\rm d}K_{\varphi}} = \frac{1}{2}
 \frac{{\rm d}^2f_1}{{\rm d}K_{\varphi}^2}\,.
\end{equation}
This function is negative around a local maximum of $f_1(K_{\varphi})$, or in
any regime where an upper bound on curvature is achieved by the modification.

Alternatively, as shown in \cite{SphSymmCov}, it is possible to construct a
partially Abelianized system with holonomy modifications, using the methods of
\cite{LoopSchwarz}. If one replaces the local Hamiltonian constraint ${\cal
  H}$ by a linear combination
\begin{equation}
 C:= \frac{(E^x)'}{E^{\varphi}}{\cal H}- 2 f_2(K_{\varphi})
 \frac{\sqrt{E^x}}{E^{\varphi}} {\cal D}
\end{equation}
with the local diffeomorphism constraint ${\cal D}$, the component $K_x$
cancels out and greatly simplifies the bracket of two constraints $C$: We have
\begin{equation}
 C=-\frac{1}{G} \frac{{\rm d}}{{\rm d}x}
   \left(\sqrt{E^x}\left(1-\left(\frac{(E^x)'}{2E^{\varphi}}\right)^2\right)+
     \frac{1}{2} \frac{(E^x)'}{\sqrt{E^x}} f_1(K_{\varphi})+
     2\sqrt{E^x}K_{\varphi}'f_2(K_{\varphi})\right)\,.
\end{equation}
One can show that the Abelianization succeeds if $C$ is a total derivative by
$x$, which requires the same condition (\ref{f2f1}) that follows from anomaly
freedom. If, on the other hand, one starts with the classical Abelianized
system and modifies the dependence of $C$ on $K_{\varphi}$, as proposed in
\cite{LoopSchwarz}, hypersurface-deformation brackets can be recovered only if
there is a modification function $\beta$ in the structure functions, implying
signature change \cite{SphSymmCov}.

However, obstructions to recovering hypersurface-deformation brackets, and
therefore obstructions to covariance, are obtained if one tries to couple a
matter field to the modified system. While this can be done within the
Abelianized system \cite{LoopSchwarz3}, it is no longer possible to recover
hypersurface-deformation brackets from the modified system
\cite{SphSymmCov}. Another question not yet addressed is whether additional
quantum effects, such as ordering choices, may affect the outcome. This
question is the topic of the following subsection.

\subsection{Operator ordering in Abelianization of the constraints}

It is possible to turn the modified Abelianized constraints into operators
without anomalies. However, this quantization step could introduce additional
properties that prevent one from recovering hypersurface-deformation brackets
even in the vacuum case. This possibility turns out to be realized for the
Schwarzschild model, and for general 2-dimensional dilaton models as well as
we will show later.

We demonstrate the sensitivity to ordering questions by working with quantized
constraints at a formal level. We only indicate the non-commuting nature of
our variables by paying attention to their position in products, but do not go
into details of regularizations such as point splitting.

We begin with the partially Abelianized system of constraints, as derived in
\cite{LoopSchwarz} but written here as operators:
\begin{eqnarray}\label{SSAbelianConstraints}
\hat{\mathcal{C}}_{\mathrm {SS}} &=& -\frac{\d}{\d x} \[\sqrt{\hat{E}^x}
\(1  - \hat{\Gamma}_\varphi^2 + \hat{f}^2\(K_\varphi\)\) + 2G M
\hat{\mathbb{I}} \]\,\\
\hat{\mathcal{D}}_{\mathrm {SS}} &=& -\frac{1}{2} (\hat{E}^x)^\prime \hat{K}_x
+ \hat{K}_\varphi^\prime \hat{E}^\varphi\,. \label{D}
\end{eqnarray}
The algebra of these constraints is such that any factor ordering choice is
allowed for $\hat{\mathcal{D}}_{\mathrm {SS}}$ while
$\hat{\mathcal{C}}_{\mathrm {SS}}$ is free from operator ordering
ambiguities. At the formal level we are working with here, it is easy to see
that two smeared $\hat{\mathcal{C}}_{\mathrm {SS}}$ commute because any
non-zero commutator introduces a delta function which gives zero once the
antisymmetric bracket is imposed.

Let us start with a factor ordering choice for the diffeomorphism constraint
as shown in Eq.~(\ref{D}). For simplicity, we will not write hats anymore, but
all expressions in this subsection remain objects sensitive to ordering
choices.  We can recover the original Hamiltonian and diffeomorphism
constraints as generators of the hypersurface deformation algebra, if we write
them as linear combinations
\begin{eqnarray}\label{SSOrgConstraints}
\mathcal{H}_{\mathrm {SS}} := \(\frac{E^\varphi}{(E^x)^\prime}\)
\mathcal{C}_{\mathrm {SS}} + 2\(\frac{f\(K_\varphi\) \sqrt{E^x}}{(E^x)^\prime}\)
\mathcal{D}_{\mathrm {SS}}\,,
\end{eqnarray}
while $\mathcal{D}_{\mathrm {SS}}$ remains unmodified. (This procedure
retraces the steps taken in \cite{LoopSchwarz}.) We begin by rewriting
$\mathcal{C}_{\mathrm {SS}}$ as
\begin{equation}
\mathcal{C}_{\mathrm {SS}} =
-\frac{(E^x)^\prime}{2\sqrt{E^x}}\(1-\Gamma_\varphi^2\) + 2\sqrt{E^x}\,
\Gamma_\varphi \Gamma_\varphi^\prime - \frac{(E^x)^\prime}{2\sqrt{E^x}}
f^2\(K_\varphi\) - 2\sqrt{E^x}\, f\(K_\varphi\) K_\varphi^\prime\,.
\end{equation}
Consequently, for our factor ordering choice for the diffeomorphism
constraint, the Hamiltonian constraint (as defined in
(\ref{SSOrgConstraints})) has the form
\begin{eqnarray}\label{SSNotQuiteH}
\mathcal{H}_{\mathrm {SS}} &=& -\frac{E^\varphi}{2\sqrt{E^x}} \(1 -
\Gamma_\varphi^2\) + 2\sqrt{E^x}\, \Gamma_\varphi^\prime -
\frac{E^\varphi}{2\sqrt{E^x}} f^2(K_\varphi) - \frac{2 E^\varphi
  \sqrt{E^x}}{(E^x)^\prime}\, f(K_\varphi) K_\varphi^\prime\nn\\
& &\qquad  -f(K_\varphi) \sqrt{E^x} K_x + \frac{2
  \sqrt{E^x}\,f(K_\varphi)}{(E^x)^\prime} K_\varphi^\prime E^\varphi\,.
\end{eqnarray}
This expression does not reduce to the usual one of the Hamiltonian constraint
because the last term in the first line of Eq.~(\ref{SSNotQuiteH}) does not
cancel out with the last term in the second line. Instead, a non-zero
commutator $[E^{\varphi},f(K_{\varphi})K_{\varphi}']$ remains.  Thus it is not
possible to obtain the generators of the hypersurface deformation algebra from
the newly defined system of constraints, once operator orderings are taken
into account.

We could have started with the other factor ordering choice in the
diffeomorphism constraint from the beginning. This is the only other choice
left to be exploited because $\mathcal{C}_{\mathrm {SS}}$ does not have any
factor order ambiguities in it. Had we started with the diffeomorphism
constraint of the form
\begin{equation}
\mathcal{D}_{\mathrm {SS}} = -\frac{1}{2} (\hat{E}^x)^\prime \hat{K}_x
+ \hat{E}^\varphi \hat{K}_\varphi^\prime \,
\end{equation}
we would have, for the Hamiltonian constraint, the expression
\begin{eqnarray}\label{SSNotQuiteH2}
\mathcal{H}_{\mathrm {SS}} &=& -\frac{E^\varphi}{2\sqrt{E^x}} \(1 -
\Gamma_\varphi^2\) + 2\sqrt{E^x}\, \Gamma_\varphi^\prime -
\frac{E^\varphi}{2\sqrt{E^x}} f^2(K_\varphi) - \frac{2 E^\varphi
  \sqrt{E^x}}{(E^x)^\prime}\, f(K_\varphi) K_\varphi^\prime\nn\\
& &\qquad  -f(K_\varphi) \sqrt{E^x} K_x + \frac{2
  \sqrt{E^x}\,f(K_\varphi)}{(E^x)^\prime} E^\varphi K_\varphi^\prime \,.
\end{eqnarray}
Once again, the unwanted terms do not cancel out. We conclude that,
with operator orderings, one cannot reproduce the hypersurface deformation
algebra from the newly defined system of constraints with a (partially)
Abelianized algebra.

\section{Dilaton gravity models}

For a general 2-dimensional dilaton gravity model with potential $V(\phi)$,
the Hamiltonian constraint in the connection variables of \cite{SphSymmPSM}
differs from the spherically symmetric one only in the term that does not
depend on extrinsic curvature or spatial derivatives of the densitized triad;
see Eq.~(46) in \cite{SphSymmPSM}, or (\ref{HDil}) here. This term does not
affect the constraint algebra, and therefore the same partial Abelianization
found in \cite{LoopSchwarz} for spherically symmetric models can be applied to
arbitrary 2-dimensional dilaton gravity models. The general Abelianized
constraints are
\begin{eqnarray}
\mathcal{C}_{\mathrm {SS}} &=& -\frac{\d}{\d x} \[\sqrt{\hat{E}^x}
\(\frac{W(E^x)}{\sqrt{E^x}}  -
\hat{\Gamma}_\varphi^2 + \hat{f}^2\(K_\varphi\)\) + 2G_2 M
\hat{\mathbb{I}} \]\,\\
\mathcal{D}_{\mathrm {SS}} &=& -\frac{1}{2} (\hat{E}^x)^\prime \hat{K}_x
+ \hat{K}_\varphi^\prime \hat{E}^\varphi
\end{eqnarray}
where ${\rm d}W(E^x)/{\rm d}x = V(E^x)$. This simple result can also be found
in \cite{AbelDil}. The same ordering obstructions as in the spherically
symmetric model follow thanks to the closely related structure of the
constraints, as do matter obstructions to be discussed in more detail in
Sec.~\ref{s:Matter}.

\subsection{Simplified kinetic term}

Using a canonical transformation, as shown in Sec.~\ref{s:Kinetic}, the
Hamiltonian and diffeomorphism constraints of 2-dimensional dilaton models can
be brought to the form
\begin{eqnarray}\label{CGHSconstraints} H[N] &=& -\int \d x
N \[4K_\varphi K_x + 4 V(E^x) E^\varphi + \frac{1}{4}\(E^x\)^\prime
\(E^\varphi\)^\prime \(E^\varphi\)^{-2}
-\frac{1}{4}\(E^x\)^{\prime\prime}\(E^\varphi\)^{-1}
\]\,,\nonumber\\
D[N^x] &=& \int \d x N^x \[E^\varphi K_\varphi^\prime -\(E^x\)^\prime K_x\]\,.
\end{eqnarray}
(The two dimensional Newton's constant $G_2$ has been set to one for
simplicity.) Classically, the hypersurface deformation algebra does not change
by a canonical transformation and is still given by
\begin{eqnarray}\label{CLassicalHDA}
\left\{D\[N^x\], D\[M^x\]\right\} &=& D\[\mathcal{L}_{N^x} M^x\]\\
\left\{H[N], D[N^x]\right\} &=& -H\[\mathcal{L}_{N^x} N\]\\
\left\{H[N_1], H[N_2]\right\} &=& D\[q^{xx}\(N_1N^\prime_2 - N_2N^\prime_1\)\]\,.
\end{eqnarray}
The only non-vanishing component of the (inverse of the) 1-dimensional spatial
metric, $q^{xx}=1/\(E^\varphi\)^{2}$, appears as a structure function. (After the
canonical transformation of Sec.~\ref{s:Kinetic}, the variables used here
should be identified with $\tilde{E}^{\varphi}= E^{\varphi}/\sqrt{E^x}$ with
$\alpha=1/2$. Therefore, $1/(E^{\varphi})^2$, dropping the tilde after the
transformation, is the same as the usual structure function
$E^x/(E^{\varphi})^2$ in spherically symmetric models.)

If one follows the common steps to use loop quantum gravity as a motivation of
a specific quantization of this model \cite{SphSymm,SphSymmHam}, one has
well-defined holonomy operators which are exponentiated versions of the
extrinsic-curvature components but are not weakly continuous in those
variables any longer. However, the holonomies corresponding to the $K_x$
variables are extended along the edge of a (one-dimensional) spin network. As
a consequence, the holonomy corrections arising from these are non-local in
nature and difficult to implement \cite{HigherSpatial}. (However, using
partial Abelianization techniques as in \cite{LoopSchwarz}, a suitable
redefinition of the constraints is possible to eliminate the $K_x$ variable.)
On the other hand, the point-wise holonomy operators, corresponding to the
$K_\varphi$ component, act at the nodes of the spin networks. These are local
modifications and can be included in our Hamiltonian constraint
\begin{equation}\label{CGHSmodconstraints}
H[N] = -\int \d x N \[4 f\(K_\varphi\) K_x + 4 V(E^x) E^\varphi
 + \frac{1}{4}\(E^x\)^\prime \(E^\varphi\)^\prime \(E^\varphi\)^{-2}
 -\frac{1}{4}\(E^x\)^{\prime\prime}\(E^\varphi\)^{-1}\]  \,.
\end{equation}
The diffeomorphism constraint remains unmodified as suggested by the simple
geometrical action on states of finite diffeomorphisms.

The hypersurface deformation brackets that do not contain structure functions
remain unaltered while the only modified bracket is
\begin{eqnarray}
\left\{H[N], H[M]\right\} &=& \int \d x \d y N(x) M(y) \left[\left\{4
    f\(K_\varphi(x)\) K_x(x), \frac{1}{4}\(E^x(y)\)^\prime \(E^\varphi(y)\)^\prime
    \(E^\varphi(y)\)^{-2} \right\} \right. \nn\\
& & \left.\hspace{1cm} + \left\{4 f\(K_\varphi(x)\) K_x(x),
    -\frac{1}{4}\(E^x(y)\)^{\prime\prime}\(E^\varphi(y)\)^{-1} \right\} +
  \(x\leftrightarrow y\)\right]\nn\\
&=& \int \d x \( NM' - N'M \) \left[-\dot{f} K_x E^x \(E^\varphi\)^{-2} - f
  \(E^\varphi\)^\prime \(E^\varphi\)^{-2} \right.\nn\\
& &\hspace{4cm} \left.+ f \(E^\varphi\)^\prime \(E^\varphi\)^{-2} + \dot{f}
  K_\varphi^\prime \(E^\varphi\)^{-1}\right]\nn\\
&=& \int \d x \( NM' - N'M \) \(\frac{\dot{f}}{\(E^\varphi\)^2}\) \(K_\varphi^\prime
E^\varphi -K_x \(E^x\)^\prime\)\nn\\
&=&  D\[\( NM' - N'M \)
\(\frac{\dot{f}}{\(E^\varphi\)^2}\)\]\label{deformedAlg} \,.
\end{eqnarray}
As expected, the dilaton potential does not contribute to the bracket.  The
structure function $q^{xx}$ is modified by the presence of $\dot{f}:=\d f/\d
K_\varphi$, which is equal to one for the classical case but not if holonomy
modifications are present. The underlying space-time symmetry is deformed for
this system as in the Schwarzschild case. However, it is not as
straightforward to conclude that this definition can give rise to signature
change: The first derivative of the modification function appears here in the
structure function, which unlike the second derivative in (\ref{beta}) does
not necessarily change sign near a local maximum. Nevertheless, the same
relationship can be established as we will do now.

\subsubsection{Partial Abelianization of the constraint algebra}

The unsmeared modified  constraints are
\begin{eqnarray}\label{unsmrdconstraints}
\mathcal{H} &=& -4 f\(K_\varphi\) K_x - 4 V(E^x) E^\varphi
 - \frac{1}{4}\(E^x\)^\prime \(E^\varphi\)^\prime \(E^\varphi\)^{-2}
 +\frac{1}{4}\(E^x\)^{\prime\prime}\(E^\varphi\)^{-1}\,,\\
\mathcal{D} &=& E^\varphi K_\varphi^\prime -\(E^x\)^\prime K_x\,.
\end{eqnarray}
We redefine the system of constraints by keeping the diffeomorphism
constraint the same while defining a new constraint as a linear combination of
the (old) Hamiltonian constraint and the diffeomorphism constraint as
\begin{eqnarray}\label{newconstraint}
\mathcal{C} &=& \frac{1}{4} \frac{\(E^x\)^\prime}{E^\varphi} \mathcal{H} -
\frac{f\(K_\varphi\)}{E^\varphi} \mathcal{D}\nn\\
&=& -\(E^x\)^\prime V(E^x) -
\(\frac{\(E^x\)^\prime}{E^\varphi}\)\[-\frac{1}{16} \(E^x\)^\prime
\(E^\varphi\)^\prime \(E^\varphi\)^{-2} +\frac{1}{16}\(E^x\)^{\prime\prime}
\(E^\varphi\)^{-1} \] - f(K_\varphi)K_\varphi^\prime\nn\\
&=& -\(E^x\)^\prime V(E^x) -
\(\frac{\(E^x\)^\prime}{E^\varphi}\)\[-\frac{1}{16} \frac{\d}{\d
  x}\(\(E^x\)^\prime \(E^\varphi\)^{-1}\)\] - f(K_\varphi)K_\varphi^\prime\nn\\
&=& -\frac{\d}{\d x}\[W(E^x) + g(K_\varphi)\] +
\frac{1}{16}\[\(E^x\)^\prime \(E^\varphi\)^{-1}\] \frac{\d}{\d x}\[\(E^x\)^\prime
\(E^\varphi\)^{-1}\]\nn\\
&=& -\frac{\d}{\d x}\[W(E^x) + g(K_\varphi) - \frac{1}{32}\(\(E^x\)^\prime
\(E^\varphi\)^{-1}\)^2 \]
\end{eqnarray}
where, as before, ${\rm d}W(E^x)/{\rm d}E^x=V(E^x)$.  We have introduced a new
function $g(K_{\varphi})$ via $f(K_\varphi)=\d g(K_\varphi)/\d
K_{\varphi}$. The smeared version of the new constraint is obtained after
integrating by parts and using $N'$ as the new smearing function:
\begin{equation}\label{Abelianconstraint1}
C[N']=\int \d x N' \[W(E^x) + g(K_\varphi) - \frac{1}{32}\(\(E^x\)^\prime
\(E^\varphi\)^{-1}\)^2 + {\mathrm {const.}}\]\,.
\end{equation}
The new constraint, so defined, is such that it commutes with itself
\begin{equation}\label{Abelianconstraint2}
\left\{C[N'], C[M']\right\} = 0\,,
\end{equation}
while its Poisson bracket with the diffeomorphism constraint is
\begin{equation}
\left\{C[N'], D[N^x]\right\} = -C\[\mathcal{L}_{N^x} N'\]\,.
\end{equation}
The Poisson bracket between two diffeomorphism constraints remains
unaltered.

The (partial) Abelianization of the constraint algebra helps us in
demonstrating signature change in all dilaton models, including the CGHS black
hole model. If one were to follow \cite{LoopSchwarz}, one would first
Abelianize the algebra classically and then insert the holonomy modification
function in the new constraint. This procedure, sometimes called
`polymerization', replaces the extrinsic curvature component $K_\varphi$ with
a bounded function of $K_{\varphi}$, usually $\sin(\rho K_\varphi)/\rho$, in
the new constraint. Claims about singularity resolution, as in
\cite{LoopSchwarz,AbelCGHS}, are based on this boundedness property.  However, as we
have seen in Eq.~(\ref{deformedAlg}), what appears in the deformation of the
structure function is $\d f/\d K_\varphi = \d^2 g/\d K_\varphi^2$, the second
derivative of the holonomy modification function. At the point of the
classical singularity, the bounded function reaches a maximum and therefore
its second derivative must be negative. We obtain hypersurface-deformation
brackets with the same sign for normal deformations as one would have in
the case of Euclidean gravity. Thus, singularity resolution in the vacuum CGHS
black hole in the framework of loop quantum gravity, just like for the vacuum
Schwarzschild black hole, cannot be divorced from signature-change.

\subsubsection{Operator ordering in the Abelianization}

For the spherically symmetric case, starting from the partially Abelianized
constraints, we have shown that none of the possible factor ordering choices
lead us to the required form of the hypersurface-deformation brackets. One can
repeat the analysis for all 2-dimensional dilaton models, including the CGHS
black hole model, following the same procedure and show that there is a
similar obstruction. Instead, we shall arrive at the same conclusion following
a different approach. We shall find a suitable operator ordering for the
generators of the hypersurface-deformation brackets, the original Hamiltonian
and diffeomorphism constraints, and then try to define the new, partially
Abelianized constraint while keeping in mind that we are now dealing with
objects sensitive to ordering choices. (Again, our treatment of operators is
formal.)

Our first task is to find a factor ordering for the original system of
constraints such that the hypersurface-deformation brackets are
realized. Following \cite{SphSymmOp}, one factor ordering choice with closed
brackets is
\begin{eqnarray}\label{factorCGHSconstraints}
\hat{H}[N] &=& -\int \d x N(x) \left[4 (\hat{E}^\varphi)^{-1} \hat{f}\(K_\varphi\)
  \hat{E}^\varphi \hat{K}_x + 4 \hat{V}(E^x) \hat{E}^\varphi
 + \frac{1}{4}(\hat{E}^x)^\prime (\hat{E}^\varphi)^\prime (\hat{E}^\varphi)^{-2}
\right.\nn\\
  & & \hspace{3cm} \left. -\frac{1}{4}
    (\hat{E}^x)^{\prime\prime}(\hat{E}^\varphi)^{-1}\right]\,,\\
\hat{D}[N^x] &=& \int \d x N^x(x) \left[\hat{K}_\varphi^\prime \hat{E}^\varphi -
  (\hat{E}^x)^\prime \hat{K}_x\right]\,.
\end{eqnarray}
The operator ordering of the Hamiltonian constraint is subtle as can be seen
from the first term in the Hamiltonian constraint, where there is no triad
component $E^\varphi$ classically, but it is important to introduce this into
the quantum operator. The resulting operator version of the
hypersurface-deformation brackets takes the form
\begin{eqnarray}\label{factorAlgebraCGHS}
\left[\hat{D}\[N^x\], \hat{D}\[M^x\]\right] &=& \hat{D}\[\mathcal{L}_{N^x}
M^x\]\\
\left[\hat{H}[N], \hat{D}[N^x]\right] &=& -\hat{H}\[\mathcal{L}_{N^x} N\]\\
\left[\hat{H}[N_1], \hat{H}[N_2]\right] &=& \hat{D}\[(\hat{E}^\varphi)^{-2}
\(\widehat{\frac{\d f}{\d K_\varphi}}\)  \(N_1N^\prime_2 - N_2N^\prime_1\)\]\,.
\end{eqnarray}
The details of the operator ordered constraints closing into an algebra are
shown in the Appendix.

The next step is to start from these constraint operators and
try to define a factor ordered new set of constraints for a (partially)
Abelianized system. In fact what we shall find below is that there is no
consistent way to carry out the Abelianization procedure anymore, once we start
from these factor ordered constraint operators. We try to define the new
constraint in such a way that we cancel out the $E^x K_x$ term coming from
the Hamiltonian constraint operator with a similar term from the
diffeomorphism constraint operator.
\begin{equation}\label{CGHSnewconstraintop}
\hat{\mathcal{C}} = \frac{1}{4} \frac{(\hat{E}^x)^\prime}{\hat{E}^\varphi}
\hat{\mathcal{H}} - (\hat{E}^\varphi)^{-2} \hat{f}\(K_\varphi\) \hat{E}^\varphi\,
\hat{\mathcal{D}}
\end{equation}
As is evident from Eq.~(\ref{CGHSnewconstraintop}), operator ambiguities play
a major role in choosing the pre-factor of the second term. This nontrivial
operator ordering is chosen so as to make the cancellation mentioned above
possible. However, the term which was a total derivative of the holonomy
correction function earlier is more complicated due to the structure of the
factor ordering chosen
\begin{eqnarray}
\hat{\mathcal{C}} &=& -\hat{V}(E^x)(\hat{E}^x)^\prime  - (\hat{E}^x)^\prime
(\hat{E}^\varphi)^{-1}\[-\frac{1}{16} (\hat{E}^x)^\prime (\hat{E}^\varphi)^\prime
(\hat{E}^\varphi)^{-2} -\frac{1}{16}(\hat{E}^x)^{\prime\prime}
(\hat{E}^\varphi)^{-1} \]\nn\\
 & & \hspace{5cm} - (\hat{E}^\varphi)^{-2}
 \hat{f}(K_\varphi)\hat{E}^\varphi\hat{K}_\varphi^\prime\hat{E}^\varphi
\end{eqnarray}
The last term in this expression is the reason why we do not have an
Abelianized system of constraints any longer.

\section{Obstructions to add matter to the holonomy-modified CGHS model}
\label{s:Matter}

In this section, we refer specifically to the CGHS model because of its
importance in discussions of Hawking radiation, which require a scalar
field. However, the same conclusions can easily be achieved for general
2-dimensional dilaton models.

If we add the simplest type of matter, a minimally coupled scalar, to the CGHS
model, the total constraints take the form
\begin{eqnarray}
H_{\mathrm {CGHS}}[N] &=& \int \d x N \left[\mathcal{H}_{\mathrm {grav}} +
  \mathcal{H}_{\mathrm {matter}}\right]\,,\\
D_{\mathrm {CGHS}}[N^x] &=& \int \d x N^x \left[\mathcal{D}_{\mathrm {grav}} +
  \mathcal{D}_{\mathrm {matter}}\right]\,,
\end{eqnarray}
with the gravitational and matter parts of the constraints given by
\begin{eqnarray}
\mathcal{H}_{\mathrm {grav}} &=& -4 K_\varphi K_x - 4 \lambda^2 E^\varphi
 - \frac{1}{4}\(E^x\)^\prime \(E^\varphi\)^\prime \(E^\varphi\)^{-2}
 +\frac{1}{4}\(E^x\)^{\prime\prime}\(E^\varphi\)^{-1}\,,\\
\mathcal{H}_{\mathrm {matter}} &=& \frac{1}{4}P_\varphi^2 \(E^\varphi\)^{-1} +
\(\varphi^\prime\)^2 \(E^\varphi\)^{-1}\,,\\
\mathcal{D}_{\mathrm {grav}} &=& -K_x (E^x)^\prime + E^\varphi K_\varphi^\prime\,,\\
\mathcal{D}_{\mathrm {matter}} &=& P_\varphi \varphi^\prime\,.
\end{eqnarray}
We have extended the gravitational phase space by a scalar field, satisfying
the Poisson bracket $\{ \varphi(x) , P_\varphi(y)\} = \delta(x,y)$. The
classical total constraints satisfy the usual hypersurface-deformation
brackets, as expected.

However, once we incorporate holonomy effects in the constraints,
$\mathcal{H}_{\mathrm {grav}}$ contains a modification function $\( K_\varphi
\rightarrow f\(K_\varphi\)\)$ whereas $\mathcal{H}_{\mathrm {matter}}$ does not
since there are no $K_\varphi$ components in the matter part of the Hamiltonian
constraint. Thus, we have
\begin{eqnarray}
\left\{H_{\mathrm {grav}}[N_1] , H_{\mathrm {grav}}[N_2]\right\} &=&
D_{\mathrm {grav}}\[\( NM' - N'M \)
\(\frac{\dot{f}\(K_\varphi\)}{\(E^\varphi\)^2}\)\]\,,\\
\left\{H_{\mathrm {matter}}[N_1] , H_{\mathrm {matter}}[N_2]\right\} &=&
D_{\mathrm {matter}}\[\( NM' - N'M \) \(\frac{1}{\(E^\varphi\)^2}\)\]\,,
\end{eqnarray}
with the gravitational and matter parts of the Hamiltonian constraint now
satisfying different versions of covariance. The Poisson bracket of the total
Hamiltonian constraint with itself does not close into the full
diffeomorphism constraint and one does not have a first-class system any
longer. One natural recourse might be to introduce a holonomy correction
function in the matter part of the Hamiltonian constraint, by hand, but then
we have a non-vanishing cross term between the gravitational and matter parts,
\begin{equation}
\left\{H_{\mathrm {grav}}[N_1] , H_{\mathrm {matter}}[N_2]\right\} -(N_1
\leftrightarrow N_2) \neq 0\,,
\end{equation}
which still leads to anomalies. For details, see \cite{SphSymmCov}.

Another way to address this problem would be to try and (partially) Abelianize
the full constrained system (with both gravity and matter), as we did
before. In this case, for the classical constraints, the Abelianization goes
through due to some subtle cancellations as shown below. That the classical
constraint algebra remains Abelianized has already been shown in
\cite{AbelDil}. We show crucial parts of the calculation here which help us to
emphasize why the quantum algebra is not anomaly free.
We write
\begin{eqnarray}
\mathcal{C}_{\mathrm {CGHS}} &=& \frac{1}{4} \frac{\(E^x\)^\prime}{E^\varphi}
\mathcal{H} - \frac{f\(K_\varphi\)}{E^\varphi} \mathcal{D}\nn\\
&=& \mathcal{C}_{\mathrm {grav}} + \mathcal{C}_{\mathrm {matter}}\,,
\end{eqnarray}
where we have defined
\begin{eqnarray}
\mathcal{C}_{\mathrm {grav}} & := & -\lambda (E^x)^\prime - K_\varphi
K_\varphi^\prime -
\frac{1}{16}\(\frac{(E^x)^\prime}{E^\varphi}\)\[(E^x)^{\prime\prime} -
\frac{(E^x)^\prime (E^\varphi)^\prime}{(E^\varphi)^2}\]\,,\\
\mathcal{C}_{\mathrm {matter}} & := & -\frac{1}{16}
\frac{(E^x)^\prime}{(E^\varphi)^2} P_\varphi^2 - \frac{1}{4}
\frac{(E^x)^\prime}{(E^\varphi)^2} (\varphi^\prime)^2 - \frac{K_\varphi}{E^\varphi}
P_\varphi \varphi^\prime\,.
\end{eqnarray}
The Poisson bracket $\left\{C_{\mathrm {CGHS}}[N_1] , C_{\mathrm
    {CGHS}}[N_2]\right\}$ can be decomposed as
\begin{eqnarray}
\left\{C_{\mathrm {CGHS}}[N_1] , C_{\mathrm {CGHS}}[N_2]\right\} &=&
\left\{C_{\mathrm {grav}}[N_1] , C_{\mathrm {grav}}[N_2]\right\}
+\left\{C_{\mathrm {matter}}[N_1] , C_{\mathrm {matter}}[N_2]\right\} \nn\\
&=& \left\{C_{\mathrm {grav}}[N_1] , C_{\mathrm {matter}}[N_2]\right\} - (N_1
\leftrightarrow N_2)\,.
\end{eqnarray}
We already know from Eqs.~(\ref{Abelianconstraint1}) and
(\ref{Abelianconstraint2}) that $\left\{C_{\mathrm {grav}}[N_1] ,
  C_{\mathrm {grav}}[N_2]\right\} = 0$. (Although we write the gravitational
part of the new constraint $C[N]$ in a different way, it is essentially the
same as the total derivative term in (\ref{newconstraint}) with the
modification function equal to the classical one.) If we calculate the other
brackets, we find that both $\left\{C_{\mathrm {matter}}[N_1] , C_{\mathrm
    {matter}}[N_2]\right\}$ and $\left\{C_{\mathrm {grav}}[N_1] , C_{\mathrm
    {matter}}[N_2]\right\} - (N_1 \leftrightarrow N_2)$ are non-zero but
they cancel each other. But the most crucial cancellation from our point of
view is the following, where the Poisson bracket proportional to
\begin{equation}
\left\{-\frac{1}{16} \frac{(E^x)^\prime}{(E^\varphi)^2} P_\varphi^2\; , \;
  -\frac{1}{4} \frac{(E^x)^\prime}{(E^\varphi)^2} (\varphi^\prime)^2\right\}
\end{equation}
from $\left\{C_{\mathrm {matter}}[N_1] , C_{\mathrm {matter}}[N_2]\right\}$ is
cancelled by terms proportional to
\begin{equation}
\left\{\frac{1}{16}\(\frac{(E^x)^\prime}{E^\varphi}\)^2
  \frac{(E^\varphi)^\prime}{(E^\varphi)}\; ,\; - \frac{K_\varphi}{E^\varphi} P_\varphi
  \varphi^\prime \right\}
\end{equation}
coming from $\left\{C_{\mathrm {grav}}[N_1] , C_{\mathrm
    {matter}}[N_2]\right\} - (N_1 \leftrightarrow N_2)$. This cancellation can
take place precisely due to the linear factor of $K_\varphi$ in the last term
of the matter part of the new constraint. If this factor is replaced by the
holonomy modification function as we should have to in case of holonomy
modifications, these two terms do not cancel anymore and we, once again, end
up with an anomaly. This is just another way of expressing that we cannot add
a massless scalar to the holonomy-modified CGHS model in a covariant
quantization, at least in the standard regularization procedure of loop
quantum gravity.

Our result extends a set of no-go theorems which had previously been proved
for the Schwarzschild black hole \cite{SphSymmCov} and Gowdy models
\cite{GowdyCov}, to the CGHS black hole and all other 2-dimensional dilaton
models. An open question was left in \cite{AbelDil,AbelCGHS} whether one could
add matter to the vacuum CGHS model and still have an anomaly free quantum
algebra. We answer this question in the negative in this article and, as a
consequence, provide an obstruction to studying Hawking radiation in this
context. It is possible to show that this obstruction can be generalized to
other possible matter models, but this calculation is essentially the same as
what has already been shown for the spherically symmetry case in
\cite{SphSymmCov}.

\section{Conclusions}

We have obtained new obstructions to anomaly-free midisuperspace models with
holonomy modifications, given by ordering effects and matter terms. The second
set of obstructions is particularly important because it spoils attempts to
discuss Hawking radiation. Such a discussion is possible within a partially
Abelianized system, only when one considers a \textit{background treatment}, but such models cannot be covariant\footnote{Even for such partially Abelianized systems, one has obstructions to adding matter to the effective theory in a covariant manner \cite{SphSymmCov}.}. In cases in which anomaly freedom can be achieved, holonomy modifications are accompanied by signature change. According to \cite{Normal}, new non-classical space-time structures are then obtained.

\section*{Acknowledgements}

This work was supported in part by NSF grant PHY-1607414.

\begin{appendix}

\section{Factor ordering for the CGHS constraints}

We introduced an operator ordering for the gravitational CGHS constraints in
(\ref{factorCGHSconstraints}) which gives a closed form of the constraint
brackets (\ref{factorAlgebraCGHS}). It is easy to observe that the subset of
the full hypersurface-deformation brackets, involving at least one
diffeomorphism constraint, works with this factor ordering. The most subtle
calculation is for the bracket between two Hamiltonian constraint operators
which has been shown below $\[\hat{H}[N]\,,\,\hat{H}[M]\] + (N \leftrightarrow
M)$. (We drop the hats from now on.)

The bracket between the first term and the third one of $\[H[N]\,,\,H[M]\]$ is
\begin{equation}\label{HH1}
\int \mathrm{d}x\, \mathrm{d}y \,N(x)M(y)
\left[4(E^\varphi(x))^{-1}f({K_\varphi}(x))E^{\varphi}(x)K_x(x)\;,\; \frac{1}{4}
  ((E^{x\prime}(y))({E^{\varphi}}(y))^{-2}({E^{\varphi\prime}}(y)) \right] \,.
\end{equation}
Combining this with the corresponding commutator between the last term of
$H[N]$ and the first term of $H[M]$, we get
\begin{equation}\label{HH2}
\int \mathrm{d}x\, (N'M-M'N)\left\{({E^\varphi})^{-2}\right\}\left\{\dot{f}(K_x)
  E^{x\prime}K_x+ f({K_\varphi})E^{\varphi\prime}\right\}\,.
\end{equation}

The bracket between the first term and the last one in $\[H[N]\,,\,H[M]\] + (N
\leftrightarrow M)$ give the terms
\begin{equation}\label{HH3}
\int \mathrm{d}x\, \[(N'M-M'N)\left\{- 2E^{\varphi\prime}
  ({E^\varphi})^{-3}f_2({K_\varphi}){E^\varphi}\right\} +
(N''M-M''N)\left\{({E^\varphi})^{-2}f({K_\varphi}){E^\varphi}\right\}\].
\end{equation}
Performing an intergration by parts on the last term, we have
\begin{equation}\label{HH4}
-\int \mathrm{d}x\,
(N'M-M'N)\left\{({E^\varphi})^{-2}\dot{f}({K_\varphi})K^\prime_\varphi {E^\varphi}
  +({E^\varphi})^{-2}f({K_\varphi})E^{\varphi\prime}\right\}\,.
\end{equation}

The required cancellation between the second term of (\ref{HH4}) and the
second term of (\ref{HH2}), as it happens in the classical case, is obtained
here with our choice of the factor ordering for the Hamiltonian constraint:
\begin{eqnarray}
\left[H[N],H[M]\right]
&=&\int\mathrm{d}x\,\left[N(x)M'(x)-N'(x)M(x)\right]
\left\{({E^\varphi}(x))^{-2}\left(
    \frac{\mathrm{d}f(K_\varphi)}{\mathrm{d}K_\varphi}\right)\right\}\nn\\
&  & \hspace{4cm}\left\{K'_\varphi(x) {E^\varphi}(x)- E^{x\prime}(x) K_x(x)\right\}.
\end{eqnarray}
The term on the right-hand side is the required diffeomorphism constraint,
with the proper operator ordering, as proposed in
(\ref{factorCGHSconstraints}).

\end{appendix}


\begin{thebibliography}{10}

\bibitem{DeformedRel}
M.\ Bojowald and G.~M.\ Paily,
\newblock Deformed General Relativity,
\newblock {\em Phys.\ Rev.\ D} 87 (2013) 044044, [arXiv:1212.4773]

\bibitem{SphSymmComplex}
J.\ Ben~Achour, S.\ Brahma, and A.\ Marciano,
\newblock Spherically symmetric sector of self dual Ashtekar gravity coupled to
  matter: Anomaly-free algebra of constraints with holonomy corrections,
  [arXiv:1608.07314]

\bibitem{CosmoComplex}
J.\ Ben~Achour, S.\ Brahma, J.\ Grain, and A.\ Marciano,
\newblock A new look at scalar perturbations in loop quantum cosmology:
  (un)deformed algebra approach using self dual variables, [arXiv:1610.07467]

\bibitem{JR}
J.~D.\ Reyes,
\newblock {\em Spherically Symmetric Loop Quantum Gravity: Connections to
  2-Dimensional Models and Applications to Gravitational Collapse},
\newblock PhD thesis, The Pennsylvania State University, 2009

\bibitem{LTBII}
M.\ Bojowald, J.~D.\ Reyes, and R.\ Tibrewala,
\newblock Non-marginal LTB-like models with inverse triad corrections from loop
  quantum gravity,
\newblock {\em Phys.\ Rev.\ D} 80 (2009) 084002, [arXiv:0906.4767]

\bibitem{SphSymmCov}
M.\ Bojowald, S.\ Brahma, and J.~D.\ Reyes,
\newblock Covariance in models of loop quantum gravity: Spherical symmetry,
  [arXiv:1507.00329]

\bibitem{GowdyCov}
M.\ Bojowald and S.\ Brahma,
\newblock Covariance in models of loop quantum gravity: Gowdy systems,
  [arXiv:1507.00679]

\bibitem{SphSymmOp}
S.\ Brahma,
\newblock Spherically symmetric canonical quantum gravity,
\newblock {\em Phys.\ Rev.\ D} 91 (2015) 124003, [arXiv:1411.3661]

\bibitem{ScalarHol}
T.\ Cailleteau, J.\ Mielczarek, A.\ Barrau, and J.\ Grain,
\newblock Anomaly-free scalar perturbations with holonomy corrections in loop
  quantum cosmology,
\newblock {\em Class.\ Quant.\ Grav.} 29 (2012) 095010, [arXiv:1111.3535]

\bibitem{DeformedCosmo}
A.\ Barrau, M.\ Bojowald, G.\ Calcagni, J.\ Grain, and M.\ Kagan,
\newblock Anomaly-free cosmological perturbations in effective canonical
  quantum gravity,
\newblock {\em JCAP} 05 (2015) 051, [arXiv:1404.1018]

\bibitem{Loss}
M.\ Bojowald,
\newblock Information loss, made worse by quantum gravity,
\newblock {\em Front.\ Phys.} 3 (2015) 33, [arXiv:1409.3157]

\bibitem{Stretched}
L.\ Susskind, L.\ Thorlacius, and J.\ Uglum,
\newblock The Stretched Horizon and Black Hole Complementarity,
\newblock {\em Phys.\ Rev.\ D} 48 (1993) 3743--3761, [hep-th/9306069]

\bibitem{Normal}
M.\ Bojowald, S.\ Brahma, U.\ B\"{u}y\"{u}k\c{c}am, and F.\ D'Ambrosio,
\newblock Hypersurface-deformation algebroids and effective space-time models,
\newblock {\em Phys.\ Rev.\ D} (2016) to appear

\bibitem{BHInt}
A.\ Ashtekar and M.\ Bojowald,
\newblock Quantum Geometry and the Schwarzschild Singularity,
\newblock {\em Class.\ Quantum Grav.} 23 (2006) 391--411, [gr-qc/0509075]

\bibitem{SphSymmPSM}
M.\ Bojowald and J.~D.\ Reyes,
\newblock Dilaton Gravity, Poisson Sigma Models and Loop Quantum Gravity,
\newblock {\em Class.\ Quantum Grav.} 26 (2009) 035018, [arXiv:0810.5119]

\bibitem{Strobl}
T.\ Strobl,
\newblock Gravity in Two Spacetime Dimensions, [hep-th/0011240]

\bibitem{Ikeda}
N.\ Ikeda,
\newblock Two-Dimensional Gravity and Nonlinear Gauge Theory,
\newblock {\em Ann.\ Phys.} 235 (1994) 435--464, [hep-th/9312059]

\bibitem{PSM}
P.\ Schaller and T.\ Strobl,
\newblock Poisson Structure Induced (Topological) Field Theories,
\newblock {\em Mod.\ Phys.\ Lett.\ A} 9 (1994) 3129--3136, [hep-th/9405110]

\bibitem{SphSymm}
M.\ Bojowald,
\newblock Spherically Symmetric Quantum Geometry: States and Basic Operators,
\newblock {\em Class.\ Quantum Grav.} 21 (2004) 3733--3753, [gr-qc/0407017]

\bibitem{CGHS}
C.\ Callan, S.\ Giddings, J.\ Harvey, and A.\ Strominger,
\newblock Evanescent Black Holes,
\newblock {\em Phys.\ Rev.\ D} 45 (1992) 1005--1009, [hep-th/9111056]

\bibitem{NewVar11}
R.\ Gambini, J.\ Pullin, and S.\ Rastgoo,
\newblock New variables for 1+1 dimensional gravity,
\newblock {\em Class.\ Quant.\ Grav.} 27 (2010) 025002, [arXiv:0909.0459]

\bibitem{LoopSchwarz}
R.\ Gambini and J.\ Pullin,
\newblock Loop quantization of the Schwarzschild black hole,
\newblock {\em Phys.\ Rev.\ Lett.} 110 (2013) 211301, [arXiv:1302.5265]

\bibitem{LoopSchwarz3}
R.\ Gambini, J.\ Olmedo, and J.\ Pullin,
\newblock Quantum black holes in Loop Quantum Gravity, [arXiv:1310.5996]

\bibitem{AbelDil}
A.\ Corichi, A.\ Karami, S.\ Rastgoo, and T.\ Vuka\v{s}inac,
\newblock Constraint Lie algebra and local physical Hamiltonian for a generic
  2D dilatonic model,
\newblock {\em Class.\ Quantum Grav.} 33 (2016) 035011, [arXiv:1508.03036]

\bibitem{SphSymmHam}
M.\ Bojowald and R.\ Swiderski,
\newblock Spherically Symmetric Quantum Geometry: Hamiltonian Constraint,
\newblock {\em Class.\ Quantum Grav.} 23 (2006) 2129--2154, [gr-qc/0511108]

\bibitem{HigherSpatial}
M.\ Bojowald, G.~M.\ Paily, and J.~D.\ Reyes,
\newblock Discreteness corrections and higher spatial derivatives in effective
  canonical quantum gravity,
\newblock {\em Phys.\ Rev.\ D} 90 (2014) 025025, [arXiv:1402.5130]

\bibitem{AbelCGHS}
A.\ Corichi, J.\ Olmedo, and S.\ Rastgoo,
\newblock Vacuum CGHS in loop quantum gravity and singularity resolution,
  [arXiv:1608.06246]

\end{thebibliography}

\end{document}